\documentclass[showpacs,superscriptaddress,twocolumn,amssymb,prb,aps]{revtex4}
\usepackage{graphicx,epsfig,verbatim,SIunits,color}
\usepackage{lmodern}
\newcommand{\dC}{$^{\circ}$C}

\begin{document}
\bibliographystyle{apsrev}
\title{Capillary Condensation, Freezing, and Melting in Silica Nanopores: A Sorption Isotherm and Scanning Calorimetry Study on Nitrogen in Mesoporous SBA-15}
\author{Sebastian T. Moerz}
\affiliation{Experimentelle Physik, Universit\"at des Saarlandes, D-66041 Saarbr\"ucken, Germany}
\author{Klaus Knorr}
\affiliation{Experimentelle Physik, Universit\"at des Saarlandes, D-66041 Saarbr\"ucken, Germany}
\author{Patrick Huber}
\affiliation{Experimentelle Physik, Universit\"at des Saarlandes, D-66041 Saarbr\"ucken, Germany}
\affiliation{Departamento de Física, Pontificia Universidad Católica, Santiago, Chile}

\date{\today}

\begin{abstract}
Condensation, melting and freezing of nitrogen in a powder of mesoporous silica grains (SBA-15) has been studied by combined volumetric sorption isotherm and scanning calorimetry measurements. Within the mean field model of Saam and Cole for vapor condensation in cylindrical pores a liquid nitrogen sorption isotherm is well described by a bimodal pore radius distribution. It encompasses a narrow peak centered at 3.3~nm, typical of tubular mesopores, and a significantly broader peak characteristic of micropores, located at 1~nm. The material condensed in the micropores as well as the first two adsorbed monolayers in the mesopores do not exhibit any caloric anomaly. The solidification and melting transformation affects only the pore condensate beyond approx. the second monolayer of the mesopores. Here, interfacial melting leads to a single peak in the specific heat measurements. Homogeneous and heterogeneous freezing along with a delayering transition for partial fillings of the mesopores result in a caloric freezing anomaly similarly complex and dependent on the thermal history as has been observed for argon in SBA-15. The axial propagation of the crystallization in pore space is more effective in the case of nitrogen than previously observed for argon, which we attribute to differences in the crystalline textures of the pore solids.
\end{abstract}


\maketitle
\section{Introduction}
The physical properties of molecular assemblies spatially confined in pores a few nanometers across are still attracting increasing interest both in applied and fundamental sciences. It originates in the technological relevance of such systems in adsorption, filtering, catalysis \cite{Schueth2002}, as hybrid materials \cite{Hoffmann2006}, as well as in the usage of tailored mesoporous solids in order to template nano-structured materials \cite{Thomas2008}. Moreover these systems allow one to explore theoretical concepts and to perform thorough comparisons between simulations of nano-confined condensed matter for equilibrium \cite{Gelb1999, Schoen2007, Binder2008, Schoen2010} as well as non-equilibrium, most prominently transport processes at the nano-scale \cite{Eijkel2005, Schoch2008, Gruener2008, Kusmin2010, Gruener2011}. The advent of tailorable porous systems with ordered pore structures (when compared to disordered porous solids like Vycor) has additionally stimulated this subfield of nanoscience and -technology \cite{Davis2002}.

In terms of the thermodynamics of pore condensates two phase transitions have been addressed quite intensively over the last two decades both in theory \cite{Gelb1999, Binder2008} and in experiment \cite{Knorr2008}. Firstly, the capillary condensation transition, that is the analogon of the bulk liquid-vapor transition in spatial confinement, where the liquid forms at vapor pressures well below the saturated bulk vapor pressure. Secondly, the liquid-solid transition, which turned out to be substantially shifted in temperature and modified in its characteristics \cite{Christenson2001, Alba-Simionesco2006, Knorr2008}. In fact, both phase transformations share remarkable similarities in pore condensates, e.g. the occurrence of pronounced hysteresis and network effects in inhomogeneous pore networks. These similarities originate in their discontinuous, first-order character and the analogous physics involved, in which one phase, the liquid or solid, replaces a second phase, the vapor or liquid, respectively. This is associated with homogeneous and heterogeneous nucleation phenomena, the replacement of one, non-wetting phase by a second, wetting phase and the intimately related formation of phase boundaries and their propagation in a highly spatially confined, interface-dominated geometry \cite{Beurroies2004, Knorr2008}.

The study of these transformations is not only of pure academic interest. They are both sensitive to the geometry of the pore space, in particular to the pore diameter. Therefore, they are widely employed in order to extract pore size distributions of meso- and nanoporous materials \cite{Mitchell2008, Petrov2009}. It has turned out, however, that the aforementioned complex transformation phenomenologies hamper simple interpretations of such studies, even for the case of an array of parallel, non-interconnected tubular pores \cite{Riikonen2011}, one of the most simple restricted geometries imaginable.

Here, we present an experimental study on the condensation of nitrogen in a template-grown, mesoporous matrix (SBA-15) which is characterized by a two-dimensional hexagonally arranged array of parallel-aligned tubular channels. We analyze this data with a mean field model for vapor condensation in cylindrical pores. The resulting conclusions with regard to nitrogen condensation in pore space set the stage for a rigorous exploration of freezing and melting of the pore condensate as a function of fractional fillings as well as cooling and heating history. The study is motivated and complemented by a comparison with previously gained data sets on freezing and melting of other systems dominated by van-der-Waals interaction, i.e. argon and carbon monoxide in SBA-15 \cite{Schaefer2008, Kityk2008b}. There, a remarkable complex behavior, involving a delayering transition, heterogeneous and homogeneous melting as well as freezing contributions has been observed. The comparison between argon (Ar) and nitrogen (N$_2$) will allow us to gain new insights regarding the influence of the crystalline symmetry or stacking sequence in the case of spherical building blocks on the propagation of the crystallization front in pore space, in particular upon heterogeneous freezing.

\section{Experimental}
\subsection{Sample preparation}
Mesoporous SBA-15 was prepared according to the original recipe presented by Zhao et al. \cite{Zhao1998}. We heated 240~g of water to 35 \dC~and mixed it with 23.9~g sulfuric acid, 8.6~g tetraethyl orthosilicate and 4~g of the amphiphilic triblock copolymer PEO$_{17}$-PPO$_{54}$-PEO$_{17}$ (Pluronic 103, BASF). After stirring for five hours the mixture was heated to 105~\dC~and kept at this temperature for 24 hours. In doing so, SiO$_2$ accumulates around the polymer micelles and precipitates as fine powder. We removed the supernate by filtering and subsequently calcinated the dried precipitate at 550 \dC~to remove the polymer from the silica matrix. A small angle X-ray diffraction pattern recorded at a rotating anode exhibited the five Bragg peaks typical of the hexagonal arrangement of the main channels with a hexagonal lattice parameter $a_h=10.71\pm0.08$~nm \cite{Hofmann2005}.  
 
\subsection{Volumetric sorption isotherm and specific heat measurements}
The SBA-15 is filled into a thin-walled copper sample cell. Additionally, we add some silver wire to the SBA-15 to enhance the heat conductivity between the silica grains and thus to decrease the time the system needs to reach thermal equilibrium. The sample can be filled in-situ with nitrogen  by controlled adsorption from the vapor phase via an external, all-metal gas handling system. This method allows us to precisely control the amount of adsorbed gas and thus the filling fraction.
The copper cell itself is enclosed inside an outer container which can be filled with helium to achieve a controlled thermal coupling between the container and the sample. We use 10~mbar as a typical pressure of the helium atmosphere.
This outer container is mounted on a closed-cycle helium refrigerator and kept inside a high vacuum environment. To further suppress heat leakage, the container is surrounded by a heatable radiation shield. Using an external temperature controller, we can adjust the container temperature $T_c$ between 20~K and 300~K.
If we change $T_c$ with a certain rate $dT/dt$, we expect the temperature $T_s$ of the sample to follow $T_c$ with a shift $\Delta T = \left | T_c - T_s \right |$ due to the heat conductivity of the helium atmosphere. Given that this shift is proportional to the heat flow $dQ/dt$, we obtain the specific heat $C(T) = dQ/dT = dQ/dt*dt/dT = K* \left |T_c-T_s \right |*(dT/dt)^{\rm -1}$ with an unknown prefactor $K$.
To minimize non-equilibrium effects, we choose a temperature ramp $dT/dt$ of 3~K/h and 5~K/h for complete and partial cycles, respectively, which should be sufficiently slow to avoid temperature gradients inside the sample. At these rates typical values of the temperature shift are on the order of 100~mK. Quite elaborate thermometry allowed us to work with such low cooling and heating rates which are more than one order of magnitude smaller than what is standard with commercial equipment.

Sorption isotherms were measured using the same setup at a fixed cell temperature of 77.9~K.

\section{Results and discussion}
\subsection{Condensation}
\begin{figure}[htbp]
\epsfig{file=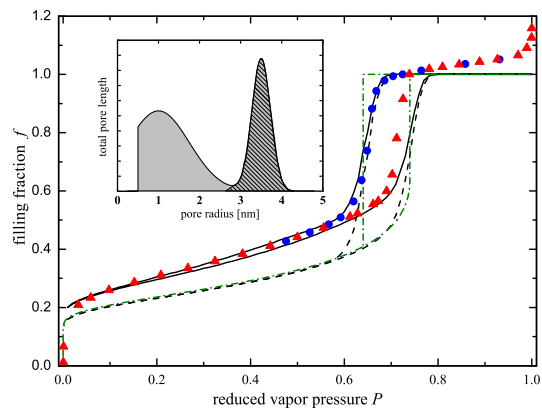, angle=0, width=0.9\columnwidth} \caption{\label{fig:N2Isotherm}(color online). N$_2$ sorption isotherm in SBA-15, measured at $77.9~\kelvin$. The dot-dashed, dashed, and solid lines represent sorption isotherms calculated based on the model of Saam and Cole for capillary condensation, while assuming a single radius, an unimodal and a bimodal Gaussian pore radius distribution (as depicted in the inset and discussed in the text), respectively.}
\end{figure}

A volumetric nitrogen sorption isotherm (normalized uptake $f=n/n_0$ as function of the reduced vapor pressure $P=p/p_0$, where $n$ is the adsorbed amount of N$_2$, $n_0=16.2$~mmol, and $p_0=1013$ mbar are the amount for complete filling of the pores and the bulk vapor pressure, resp.), recorded in the liquid regime of the N$_2$ pore filling at $77.9~$K, is shown in Fig.~\ref{fig:N2Isotherm}. 

One can qualitatively distinguish between five regimes in the isotherm, presented in the following as a function of rising $P$: The initial steep increase is due to the formation of a strongly silica-bound first monolayer of molecules and, as shall be discussed later, also related to condensation in micropores. For 0.05$\leq$P$\leq$0.6 the film grows in thickness beyond the monolayer. The reversible shape of the isotherm with respect to ad- and desorption is typical of multilayer film growth on a heterogeneous substrate. It is followed by a hysteretic regime characteristic of the filling of the pores by capillary condensation on $P$-increase (starting at about $f_{\rm A}=0.6$) and to the evaporation of the capillary bridges on $P$-decrease (terminating at about $f_{\rm D}=0.5$). The liquid bridges formed upon capillary condensation are terminated by concave menisci having curvature radii on the order of the mean pore diameter $r_0$ of the mesopores. The final slope in the post-filling regime $f>1$, when the tubular pore space is entirely filled, demonstrates that there are further sites available for condensation, with characteristic linear dimensions larger than $r_{\rm 0}$, but still finite. One can think of tapered pore mouths, niches of the rough external surface of the grains, and the points at which the powder grains are in mutual contact \cite{Wallacher2004a}. Finally, a steep increase at $P=1$ indicates the formation of bulk droplets.

In order to quantitatively analyze the condensation in SBA-15 we refer to the theory for vapor condensation in cylindrical geometry proposed by Cole and Saam \cite{Cole1974, Saam1975}. It is a mean field model for film and capillary condensation under consideration of the van-der-Waals interaction with the pore walls. The liquid-vapor surface tension $\gamma_{lv}$ and the molar volume $v_l$ enter as characteristics of the sorbate, whereas the sorbate-wall interaction is represented by a parameter $\alpha$ in respect to the fluid-fluid interaction. The dot-dashed and dashed lines in Fig. \ref{fig:N2Isotherm} represent the predictions of this theory, while assuming literature values of $\gamma_{lv}$ and a value of $\alpha=10^{-49}$Jm$^6$, determined from nitrogen adsorption in mesoporous silica for several different pore geometries: A single fixed pore diameter of $r_0$=3.3~nm, an unimodal Gaussian pore radius distribution $r_0$ with a distribution width of $\sigma=6.5$\%.  The main characteristics observed, i.e. the monolayer and multilayer film growth culminating in a hysteretic capillary condendsation regime, is correctly described by the unimodal distribution of mesopores. Note, however, that we cannot reproduce the large $f$-increase in the film-growth regime (before capillary condensation) and, simultaneously, the correct positions of the closure points of the condensation hysteresis.

Stimulated by the sizeable number of studies which highlight the existence of microporosity in the pore walls, we tried different pore radius distributions extending from the meso- to the micropore regime and calculated an isotherm by adding up Saam-Cole isotherms according to the pore radius distribution chosen. The best agreement between measurements and calculated effective isotherm could be achieved by introducing the bimodal pore radius distribution depicted as inset in Fig. \ref{fig:N2Isotherm}. It comprises mesopores with a mean pore radius of $r_{meso}=3.3$~nm and narrow width of $\sigma_{meso}=6.5$\% along with a second fraction of micropores with $r_{micro}=1.0$~nm, large distribution width $\sigma_{meso}=75$\% and a lower cut-off of $r_{min}$=0.5~nm. The hard cut-off is heuristically motivated. The microporosity originates in the removal of the templating polymer from the silica walls. The smallest resulting pore diameter should therefore represent the thickness of a single polymer chain, which is approx. 0.5~nm. 

Using those parameters we achieve a remarkable agreement between measured and calculated isotherm, in particular in the film growth regime. In fact, the Saam-Cole model predicts for this pore size distribution also a small, but final hysteresis between ad- and desorption extending down to small reduced $P$, which at least for $0.5<P<0.6$ is compatible with our measurements. Note that the deviations between measurement and theory for the metastable adsorption branch are not too surprising and can be traced to an overestimation of the metastable film thickness in the Saam-Cole model, since a perfect cylindrical geometry is considered and thermal fluctuations are neglected. Both can trigger capillary condensation and thus an adsorption branch shifted to lower $P$ than predicted by the model. Nevertheless, the Saam-Cole model yields also information on the maximum film thickness established on the pore walls upon onset of capillary condensation. For the main mesopores one obtains a values of this film thickness $t_m$ of $1.3\,$nm, corresponding to approx. 4 molecular layers of nitrogen, whereas the thickness of the film after complete capillary evaporation amounts to $t_c=0.76\,$nm corresponding to about 2-3 monolayers. Thus the thickness of the metastable film (the difference between $t_m$ and $t_c$) represents approx. 1-2 monolayers. 

Our pore size distribution attributes $28.3$\% of the overall porosity to the intrawall pore space, which is in good agreement with a value of 30\% reported from analyses of nitrogen isotherms of SBA-15 with density functional calculations \cite{Ravikovitch2001} and structural characterization by scattering techniques \cite{Findenegg2010}. Thus, along with the 10\% material condensing in the post-fillling regime in intergranular spaces, there is an overall volume fraction of 40\% nitrogen, which is not adsorbed in the main mesopores, but in micropores or in inter- and extragranular spaces.

Our results with regard to the bimodal, hierarchical pore radius distribution in SBA-15 are in good agreement with small-angle X-ray \cite{Imperor-Clerc2000, Hofmann2005, Findenegg2010, Mueter2009, Jaehnert2009, Findenegg2010, Pikus2010, Pollock2011} and neutron \cite{Pollock2011} studies as well as atomistic, lattice Monte Carlo simulations \cite{Bhattacharya2009} addressing the intrawall porosity of SBA-15. The controversially debated question, whether this pore wall porosity is corona-like, confined to a small region around the main pores, or whether it affects the entire pore walls, can not be answered by our analysis. The observation of interconnected wires in electron microscopy imaging of inverse platinum replicas of SBA-15 \cite{Ryoo2000, Galarneau2003}, which can be traced to one of the main defect structure during the templating process, corroborate rather the idea of a homogeneously distributed porosity in the silica walls. 

Over the last couple of years there has been a renewed interest in the deformation of mesoporous solids upon gas condensation \cite{Guenther2008, Schoen2010, Schoen2010a, Gor2010, Gor2011}. X-ray diffraction analyses of the pore array of SBA-15 (and other template-directed, ordered mesoporous solids, such as MCM-41) allowed for a precise investigation of the pore deformation as a function of $f$. In particular, there is a noticeable contraction of the pores upon capillary condensation. Based on the elastic properties of silica and the macroscopic approach for the adsorption-induced deformation of SBA-15 upon nitrogen condensation suggested by Gor and Neimark \cite{Gor2010, Gor2011}, this effect can be estimated to be significantly below 1\% of the pore radius, which is well below the resolution of our pore distribution analysis.
Hence we further neglect any condensation induced deformation of the porous matrix.
%


\subsection{Freezing and Melting} \subsubsection{Complete freezing/heating cycles}

\begin{figure}[htbp]
\epsfig{file=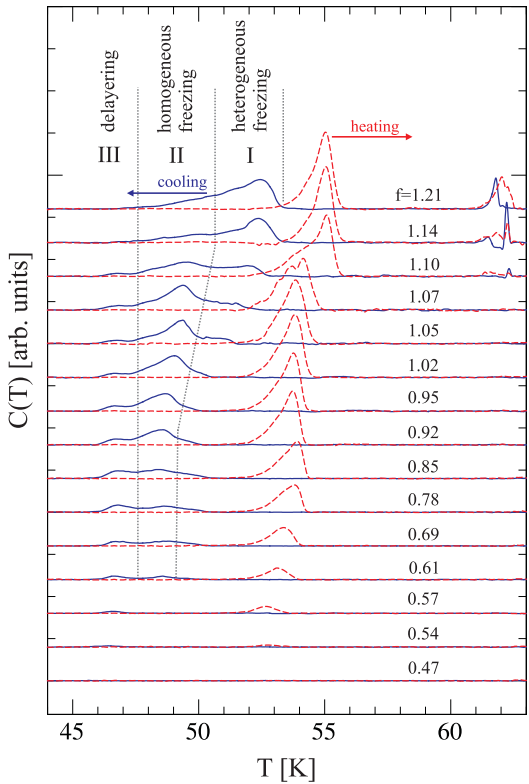, angle=0, width=0.8\columnwidth} \caption{\label{fig:N2SpecHeatFullcycles}(color online). Specific heat curves of cooling (solid line) and heating (dashed line) cycles for selected filling fractions $f$ indicated in the figure for N$_2$. For the sake of readability the curves are shifted vertically by an arbitrary offset.}
\end{figure}

In Fig. \ref{fig:N2SpecHeatFullcycles} the specific heat of condensed nitrogen measured as a function of $T$ is plotted for a selected set of $f$ for full cooling and heating cycles of the sample. Starting from temperatures above the triple point of nitrogen ($T_3^{N_2}$=63.15~K) the samples were cooled down to $T_{min}$=42~K with a cooling rate of 3~K/h. The background specific heat of the empty setup and empty porous sample ($f=0$) were subtracted from the $f>0$ measurements in order to focus solely on the caloric anomalies of the condensed nitrogen.

As it has been observed for Ar in Vycor \cite{Wallacher2001} and in SBA-15 \cite{Schaefer2008} there is no excess specific heat observable up to $f=0.47$. The wall coating up to 2-3 monolayers does not take part in a collective phase transformation. Presumably, it forms strongly bound, ''dead'' layers of short-range, triangular lattices on the heterogeneous silica wall of the main pores as it was inferred from x-ray diffraction experiments on N$_2$ condensed in Vycor \cite{Huber1998a, Huber1999a} and for N$_2$ physisorbed on aerosil silica surfaces \cite{Morishige2001}.

Starting with $f=0.54$ the filling fractions investigated exhibit both upon cooling and heating anomalies in the caloric signal. A single, asymmetric anomaly, observed in heating, can be attributed to the melting of the solidified pore condensate. The shape of the anomaly with a broad low-$T$ and a fast decaying high-$T$ wing remains unchanged for all filling fractions studied. 

\begin{figure}[htbp]
\epsfig{file=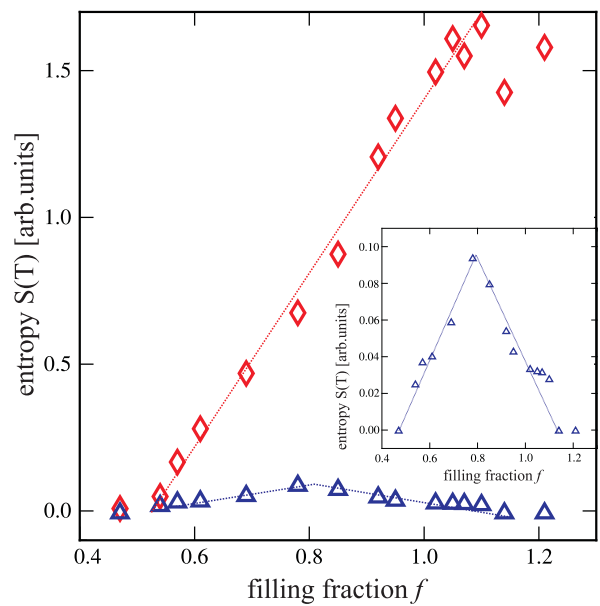, angle=0, width=0.9\columnwidth} \caption{\label{fig:N2Entropy}(color online). Melting entropy (red, solid diamonds) and entropy related to the delayering transition (anomaly III) (blue, solid triangles) as a function of filling fraction $f$. The inset depicts a zoomed version of the anomaly III entropy as a function of filling fraction $f$. The dotted lines are guides for the eye.}
\end{figure}

The  entropy of fusion, $S_{\rm 0}$, is obtained by integrating the calorimetric signal above the smooth background, $S_{\rm 0} = c \smallint(dQ'/dT)/TdT$. The coefficient $c$ has not been determined, but is the same for all heating runs. A plot of $S_{\rm 0}$ calculated from the melting curves and thus representing the melting entropy as a function of $f$ is shown in Fig.~\ref{fig:N2Entropy}. It increases linearly with $f$, indicating an $f$-independent molar entropy of fusion, and extrapolates to zero at $f_c$=0.53$\pm$0.05, close to the lower closure point of the hysteresis of the isotherm at $f_D=0.48$. This corroborates the idea that freezing and melting is restricted to the material involved in the adsorption/desorption hysteresis of the main pores, that is the capillary condensed fraction of nitrogen, and it encompasses also the material which originally formed a metastable film on the pore walls ($f_D-f_A$) \textbf{with the filling fraction $f_A$ at the onset of capillary condensation during adsorption}.

All models for solidification in pores predict, independent of their details, that the decrease of the melting/freezing temperature is proportional to the inverse mean pore radius. This simple scaling results from the competition of volume free energies (proportional to $r^3$) and excess interfacial free energies (proportional to $r^2$). Consequently it is tempting to directly relate the shape of the melting anomaly to the final pore size distribution. We prefer, however, an interpretation of this finding in terms of interfacial melting in a cylindrical pore starting at the boundary between the dead and the mobile part of the pore filling rather than by referring to a pore size distribution which had to be rather special and overall much broader than inferred from the N$_2$-isotherm in order to reproduce the peculiar shape of the anomalies.

In this model the continuous growth of the liquid boundary layer between solidified filling and pore wall gives rise to the broad low-T side of the melting anomaly. Thus, only the sharp high-T side gives a direct representation of the pore size distribution. This interfacial melting scenario is similar to surface melting at semi-infinite confined, planar surfaces \cite{Zhu1986, tenBosch1993, Engemann2004} and has extensively been outlined and discussed by Wallacher and Knorr for melting of Ar in Vycor \cite{Wallacher2001}. Calorimetry can of course not prove whether the melting front propagates along the pores or radially inward, but the mere fact that the asymmetric shape of the anomaly is recovered in the highly homogeneous pores of the present substrate is a strong argument in favor of interfacial, i.e. radial melting. \textbf{It is worth noting that the conclusions of this model do not depend on whether the filling occurs by successive and complete filling of individual pores or by parallel filling of pore segments with varying diameter, as long as the axial size of the condensate is considerably larger than the pore diameter.}

The anomalies in the calorimetric signal exhibit a much more complex behavior as a function of $f$ and $T$ upon cooling than upon heating of the pore condensate. The freezing of the condensate is spread up to 9~K from 54~K to 45~K, whereas the melting is confined in a $T$-interval of 2.5~K at maximum.

For ease of discussion we separate the cooling anomaly in three parts - I, II, and III - see Fig. \ref{fig:N2SpecHeatFullcycles}. As we shall outline in the following this partitioning can be motivated by the distinct $f$- and $T$-dependence of these parts and it can be traced to three distinct spatial arrangements of these freezing molecular populations in pore space. 

Part~III of the freezing anomaly represents the component of the filling that is the last to solidify on cooling. The related excess entropy, $S_{\rm 0}$(III) shows a peculiar type of $f$-dependence, see Fig. \ref{fig:N2Entropy}. It increases linearly with $f$ starting at $f_D=0.54$ and upon reaching $f=0.8$, the steepest point in the adsorption isotherm, it starts to decrease. Identical peculiar $f$-dependencies have been reported in specific heat measurements for Ar in Vycor \cite{Wallacher2001} as well as for water \cite{Schreiber2001} and Ar \cite{Schaefer2008} in SBA-15. As discussed in Ref. \cite{Wallacher2001} it can be attributed to the solidification of the mobile part of the liquid film on the pore walls in otherwise empty pore segments. Upon solidification the film delayers and the material involved forms (or joins already formed) solidified capillary condensate of the fractions II and I (and melts as such on subsequent heating). Thus, this fraction and the corresponding entropy scales with the mobile fraction of molecules in the vapor filled regions of pore space. This component is increasingly displaced upon onset of capillary condensation by capillary filled pore segments, which give rise to the anomalies I and II. This indirectly inferred rearrangement of pore condensate components explains not only the peculiar $f$-dependence of $S_{\rm 0}$(III). It is also in accordance with the simple single-peak melting behavior, entirely independent of $f$. Moreover, it is corroborated by $f$-dependent light scattering \cite{Soprunyuk2003} and ultrasonic measurements \cite{Schappert2011} on Ar condensed in mesoporous silica, indicating a coarsening of the pore condensate upon freezing. Further arguments supporting this delayering process will be presented in the part of the paper concerning the measurements on partial cooling/heating cycles.

Still referring to the underfilled situation, $f<1$, this reasoning leaves parts I and II for the freezing of the capillary condensate in the regular pores. We recall that the interfacial melting model presented in Ref. \cite{Wallacher2001} not only explains the lowering of the equilibrium freezing/melting temperature $T_{\rm 0}$ in pore confinement but also predicts metastable states, liquid ones down to the lower spinodal temperature $T^{\rm -}$ and solid ones up to the upper spinodal temperature $T^{\rm +}$. We propose to identify peak~II with the freezing of the capillary condensate via homogeneous nucleation of the solid phase in the pore center at $T^{\rm -}$. Peak~II is comparably sharp because of the narrow pore size distribution of SBA-15. Part~I of the freezing anomaly on the other hand stems from the freezing via heterogeneous nucleation within the rather broad ($T_{\rm 0}$, $T^{\rm -}$)-interval, triggered by the presence of various types of bulk crystallization nuclei. The chance that a parcel of liquid is in contact with such nuclei increases with $f$. That is why part~II dominates (in coexistence with part~III) at low $f$ and goes through a maximum slightly below $f=1$, whereas part~I starts to develop upon overfilling (f$\leq$1.02) and thus bulk formation, only.

The heterogeneous freezing character of part I and the importance of nucleation centers outside the main pores is particularly corroborated by the drastic decrease of the anomaly II upon overfilling of the pores ($f>1$). The solidified bulk crystallites outside the main pores induce freezing of all pore segments in direct contact with these external solidified nuclei, consequently the fraction of homogeneous freezing (part II) tends to zero upon increase of the overfilling.

If the microporosity of the walls or the larger channels inferred from micrographs of SBA-15 replicas, mentioned above, were responsible for an interconnection of the main mesopores, this could in principal increase the effectiveness of the propagation of heterogeneously nucleated crystallization fronts in pore space. However, the material in the micropores does not exhibit any freezing anomaly upon cooling, it is part of the dead molecular fraction, as the absence of any specific heat anomalies in the low-$f$ measurements demonstrate. Unfortunately, we cannot infer any direct information with regard to larger interconnecting channels.

\begin{figure}[htbp]
\epsfig{file=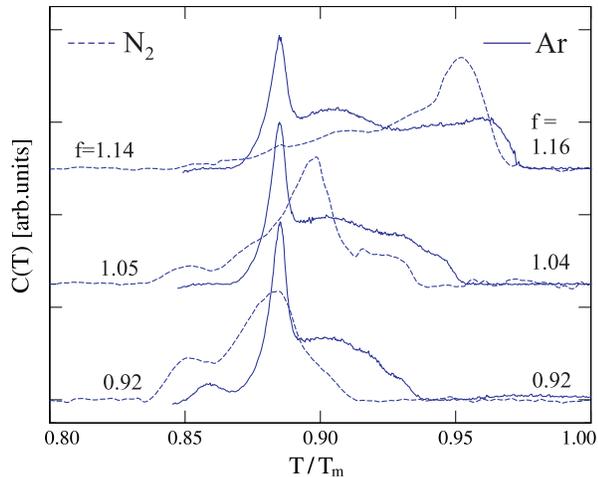, angle=0, width=0.9\columnwidth} \caption{\label{fig:ArN2Comparison}(color online). Specific heat freezing anomalies of Ar (solid lines) and $N_2$ (dashed lines) condensed in SBA-15 for selected filling fractions, as indicated in the figure, and plotted for a better comparison as a function of normalized temperature $\Theta$, where $\Theta=T/T_m$ and $T_m$ is the melting temperature of confined Ar and N$_2$, respectively, \textbf{defined by the peak of the melting anomaly in the case where the pore condensate is assumed to be in contact with bulk crystallites ($f>1.10$).}}
\end{figure}

It is interesting to compare the freezing/melting behavior of argon and nitrogen - see Fig. \ref{fig:ArN2Comparison}. In general the observed behavior for Ar is very similar to the one found for N$_2$ - a complex, broad freezing anomaly (encompassing delayering, homogeneous and heterogeneous freezing) and a narrow, one-peak melting feature. Note, however, that the vanishing of the homogeneously freezing fraction and the simultaneous increase of heterogeneous freezing is much more pronounced for N$_2$ than for Ar upon overfilling of the pores. We suggest that this difference originates in the distinct crystallographic structures of the molecular solids. Bulk Ar crystallizes in a face-centered-cubic lattice, which results from an ABC stacking of triangular net planes and which is characterized by four equivalent stacking directions along the four cubic body diagonals. By contrast, bulk N$_2$ crystallizes in a hcp lattice with a single stacking direction for the triangular net planes, the crystallographic $<001>$ direction. X-ray diffraction experiments on the crystallization of Ar and N$_2$ in mesopores revealed a conservation of these crystalline building principles, except for a substantial amount of stacking faults \cite{Huber1998, Huber1999, Huber1999a, Morishige2000, Morishige2004, Hofmann2005}. More importantly,  diffraction studies on oriented tubular pores in monolithic mesoporous silicon indicate an alignment of the stacking (and fast growing) crystalline direction(s) along the long axes of the channels \cite{Hofmann2005}, similar as it is known from the Bridgman technique of single crystal growth. 


This preferred orientation of the confined crystallites is, however, for $N_2$ with its single $<001>$ stacking direction, much more pronounced than in the case of Ar, where four ($<\pm 1 \pm 1 \pm 1>$) stacking directions along the four body diagonals compete \cite{Hofmann2005a}. With respect to the propagating crystallization front this also means that upon heterogeneous nucleation by an external nucleus the front can rapidly and more easily sweep across the entire channel in the case of N$_2$. No high cost grain boundaries have to be formed and a vanishing amount of material is left for homogeneous freezing, in agreement with our caloric experiment. By contrast in the case of Ar the nanoscale Bridgman mechanism is less effective, leaving a large part of the condensate decoupled from the externally triggered crystallization for homogeneous nucleation, presumably by arrests of the crystallization front along the channels, the formation of empty pore space due to the volume shrinkage upon solidification and mismatching grain boundaries \cite{Huber1998, Huber1999a, Hofmann2005a}. 

\begin{figure}[htbp]
\epsfig{file=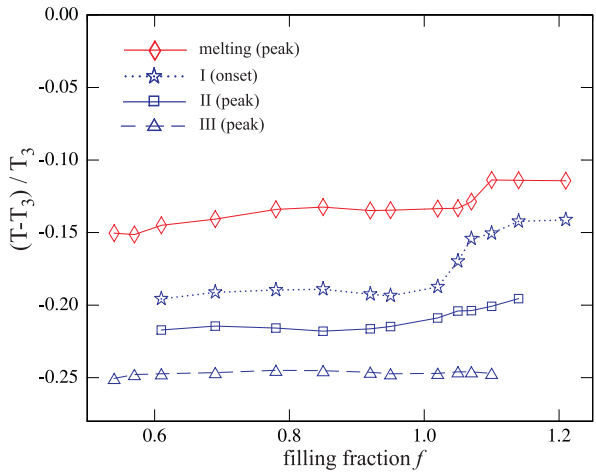, angle=0, width=0.9\columnwidth} \caption{\label{fig:N2theta}(color online). Transition temperatures T of the characteristic anomalies found in the specific heat of N$_2$ in SBA-15 as a function of filling fraction and normalized to the bulk triple point T$_3$ of N$_2$.}
\end{figure}

There is another interesting change in the position of the specific heat anomalies upon reaching $f \geq 1.1$, that is a sizeable increase both of the onset temperature of the freezing anomaly and the center of mass of the melting anomaly - compare Fig. \ref{fig:N2SpecHeatFullcycles}. In order illustrate this behavior in more detail, we plot in Fig. \ref{fig:N2theta} the characteristic temperatures of the specific heat anomalies as a function of $f$. There are no or only marginal temperature shifts for the anomalies attributed to the delayering and the homogeneously freezing component (peak II, III). By contrast the onset of heterogeneous freezing as well as the homogeneous melting is sizeably shifted towards the bulk tripel point $T_{\rm 3}$. It is tempting to relate these changes to the direct or indirect contact of these parts of the pore condensate with already or still solidified 3D material outside the pore-space achieved upon overfilling. There is, however, also an alternative or rather complementary explanation. Upon overfilling the concave meniscus of the confined liquid \cite{Kityk2008a} and solid vanishes \cite{Hoffmann2003} and intimately related, the tensile pressure acting in the pore condensate disappears \cite{Kanda2004, Morishige2006, Kanda2007} upon reaching $f=1.1$. In fact, an estimation of the tensile pressure release (of $\sim$ 54 bar) encountered by the pore condensate \cite{Schaefer2008} along with an extrapolation of the pressure-temperature melting line of N$_2$ \cite{Keesom1934} towards negative (tensile) pressures yields an upward shift of 1.2\,K for N$_2$, in excellent quantitative agreement with our observation. A similarly good quantitative agreement is achieved for Ar in SBA-15 \cite{Schaefer2008}. Also supportive for the contribution of the tensile pressure effect is the observation of an inverse, downward jump in $T$ of the freezing and melting anomaly of water in SBA-15 \cite{Schreiber2001} upon overfilling, which is expected due to the 'anomal' melting line of water. 

\subsubsection{Incomplete Freezing and Melting Cycles}

In contrast to the preceding sections discussing complete cooling-heating cycles, where the condensate had been entirely solidified before its subsequent complete melting, we present incomplete cycles in the following. We investigate the melting of partially solidified (incomplete freezing) or the freezing of incomplete liquidized samples (incomplete melting). These studies allow us to further scrutinize the melting/freezing mechanisms and the partitioning of the condensate in thermodynamically distinct fractions.

\begin{figure}[htbp]
\epsfig{file=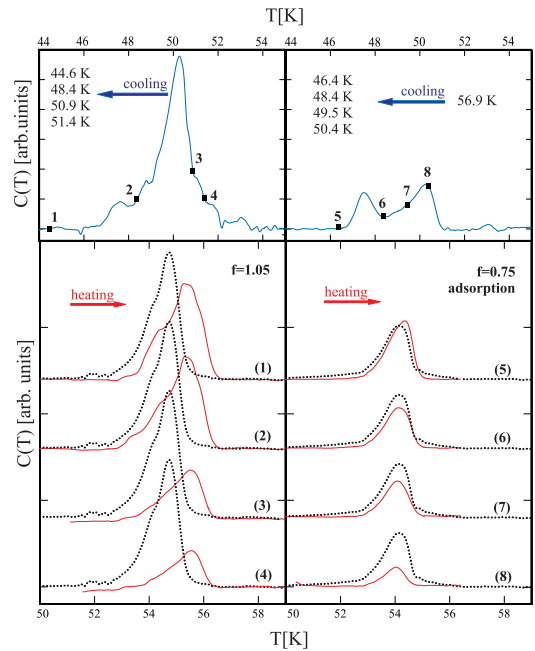, angle=0, width=0.8\columnwidth} \caption{\label{fig:IncompleteFreezing}(color online). Incomplete freezing cycles as discussed in the text. The upper panels depict the freezing anomaly of complete freezing cycles for two selected filling fractions as references. The solid symbols indicate to which temperature, listed in the figure, the sample was cooled from 56.9 K, before the heating scans, marked by numbers, started. \textbf{The dotted lines represent the melting anomaly of a complete heating run with a virgin sample.}}
\end{figure}

Starting from an entirely liquid sorbate the condensate was cooled for a partial filling ($f=0.75$) and for an overfilled sample ($f=1.05$) to four selected temperatures above and below complete freezing of the condensate - see Fig. \ref{fig:IncompleteFreezing}. Upon subsequent heating we investigate the shape of the melting anomaly and compare it with the anomaly typical of the complete cycles discussed above (dotted line).

For the partial filling the shape and position of the melting anomaly in incomplete freezing cycles is independent of the degree of solidification of the sample, except for a scaling of its area with the fraction of solidified material. In particular the delayering transition, affecting the mobile multilayers of the vapor filled pore segments, does not lead to an extra feature upon melting. However, \textbf{by comparing curves 5 and 6 in Fig. \ref{fig:IncompleteFreezing}, }the high-$T$ wing of the melting anomaly appears slightly steeper after inducing this transition (by cooling below 47~K). This can be rationalized by a coarsening of the material.

A very similar behavior is found for the overfilled sample. Independent of the degree of solidification, the melting anomaly remains a one-peak feature, albeit shifted towards higher $T$, when compared to the previous full cycle measurement of the virgin sample. The pore condensate is already coarsened by the many heating/cooling and thus solidification/melting cycles \cite{Soprunyuk2003} and the concave menisci disappeared to large extent, which results in a loss of the tensile pressure and consequently in an upward shift of the melting peak. Interestingly enough, for melting scans starting below 49~K the low-$T$ side is getting more pronounced. This could be related to the occurrence of delayering and the formation of solid capillary bridges with concave menisci (and thus reduced $T_m$) in those pore segments, which are vacant (except for multilayers) after the volume shrinkage in the pore condensate induced by the freezing of the capillary filled pore segments.   

\begin{figure}[htbp]
\epsfig{file=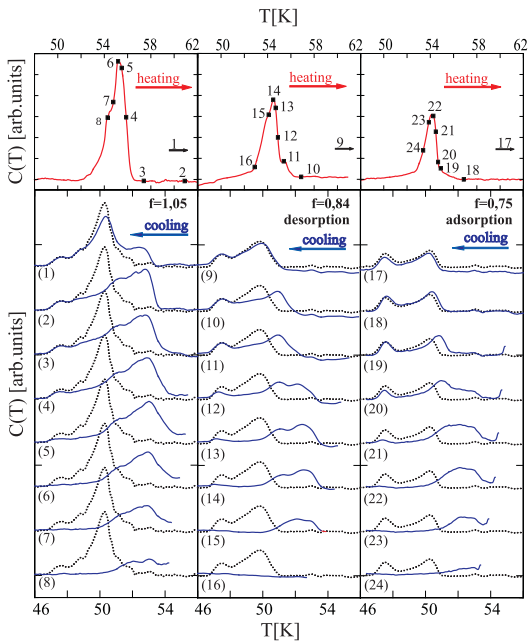, angle=0, width=0.8\columnwidth} \caption{\label{fig:IncompleteMelting}(color online). Incomplete melting cycles as discussed in the text. The upper panels depict the melting anomaly of complete freezing cycles for three selected filling fractions as references. The solid symbols indicate to which temperature, listed in the figure, the sample was heated, before the cooling scans, indicated by numbers, started.}
\end{figure}

The freezing of partially melted samples were investigated for three selected filling fractions (f=1.05, 0.84, 0.75) and for a set of eight temperatures ranging from partial to complete melting of the samples - see Fig. \ref{fig:IncompleteMelting}.

The freezing anomalies exhibit for all filling fractions investigated a quite similar behavior. Starting from only slightly molten states, in which a significant part of the sample is still in the solid phase, a fundamental change in the freezing behavior is observable, where the sorbate starts to freeze upon cooling at temperatures far above the ones observed for freezing in the full cycles. 

Upon heating of the sorbate beyond the maximum of the melting anomaly, the freezing anomaly known from the full cycles, starts to get established. First the delayering anomaly reappears, which supports the conclusion of a reversible formation of the higher adsorbate layers. Moreover a peak reminiscent of anomaly II starts to reappear at the expense of the high-$T$ freezing components. By an increase of the melted material, an increasing amount of condensate is decoupled from heterogeneous nucleation centers. The homogeneous melting is shifted towards larger $T$, when compared to the full cycles. This shift is more pronounced in the $f=0.84$ sample, prepared by desorption, than in the $f=0.75$ sample, prepared by adsorption. An observation which can be attributed to the distinct geometric arrangements of the confined liquid after these preparation procedures. Upon adsorption the liquid condenses finely dispersed in pore space, whereas upon desorption the partitioning is much more coarse with large filled and large empty segments \cite{Soprunyuk2003, Naumov2008, Kityk2009}. The spatial coarsening leads to an upward shift in $T$, since it favors the propagation of crystallization fronts.
 

\section{Conclusions}
We presented a combined calorimetric and sorption study on nitrogen confined in mesoporous SBA-15. An analysis of the condensation with the mean-field model of Saam and Cole resulted in a bimodal pore size distribution encompassing mesopores and a sizeable fraction (of about 30\%) micropores, presumably originating in the templating process.

Filling fraction dependent caloric experiments revealed temperature anomalies upon cooling spread over a broad temperature range, which can be attributed to a delayering transition, homogeneous and heterogeneous freezing. By contrast, the melting is comparably sharp in temperature. Upon freezing a rearrangement of the material both in radial and axial direction occurs. In general, our findings are qualitatively identical to observations made previously for water, carbon monoxide and nitrogen in SBA-15. The heterogeneous freezing upon overfilling of the pores (at the cost of homogeneous freezing) is much more pronounced in the case of N$_2$ than in the case of Ar, which we attribute, motivated by previous structural studies on the freezing of these molecular species in tubular pores, to the different structure (texture) of the low-temperature, crystalline phases. The simple hcp structure of N$_2$ (with one stacking direction) allows for a more effective propagation of externally induced crystallization fronts along the tubular channels. 

Our study corroborates previous findings regarding the importance of the axial propagation of the crystallization front upon pore freezing \cite{Khokhlov2007}, which render the liquid-solid transition similarly susceptible to network effects as the capillary condensation transition. Additionally it demonstrates that the effectivity of this propagation depends not only on the pore geometry. It is also sensitive on the structure of the crystalline phase to be formed. Given the growing evidence of textured crystallization of molecular building blocks, encompassing not only the spherical systems presented here, but also rod-like (n-alkanes \cite{Henschel2007, Henschel2009},  n-alcohols \cite{Henschel2008, Berwanger2009}, liquid crystalline \cite{Chahine2010}) as well as polymeric molecular species \cite{Steinhart2006, Steinhart2008, Duran2011}, we believe that our finding is of particular interest with regard to the effective control of crystalline orientation in templating matter by nanoporous templates \cite{Thomas2008}. 

We envision further studies on how solidification fronts spread as a function of crystalline symmetry and pore geometry. Computer simulations could help to get complementary, microscopic insights in this complex phenomenology \cite{Hoffmann2003, Ma2008, Teeffelen2009, Schilling2010, Wilms2010, Sandomirski2011}.



\begin{acknowledgments}
This work has been supported by the DFG graduate school 1276, `Structure formation and transport in complex systems' (Saarbr\"{u}cken).
\end{acknowledgments}




\end{document}